\documentclass[twocolumn]{el-author}

\newcommand{\ua}{\uparrow}
\newcommand{\nc}{\newcommand}
\nc{\da}{\downarrow} \nc{\hc}{\hat{c}} \nc{\hS}{\hat{S}}
\nc{\bra}{\langle} \nc{\ket}{\rangle} \nc{\eq}{equation (\ref}
\nc{\h}{\hat} \nc{\hT}{\h{T}}\nc{\be}{\begin{eqnarray}}
\nc{\ee}{\end{eqnarray}}\nc{\rd}{\textrm{d}}\nc{\e}{eqnarray}\nc{\hR}{\hat{R}}\nc{\Tr}{\mathrm{Tr}}
\nc{\tS}{\tilde{S}}\nc{\tr}{\mathrm{tr}}\nc{\8}{\infty}\nc{\lgs}{\bra\ua,\phi|}\nc{\rgs}{|\ua,\phi\ket}
\nc{\hU}{\hat{U}}\nc{\lfs}{\bra\phi|}\nc{\rfs}{|\phi\ket}\nc{\hZ}{\hat{Z}}\nc{\hd}{\hat{d}}\nc{\mD}{\mathcal{D}}
\nc{\bd}{\bar{d}}\nc{\bc}{\bar{c}}\nc{\mc}{\mathcal}\nc{\ea}{eqnarray}\nc{\mG}{\mathcal{G}}\nc{\bce}{\begin{center}}
\nc{\ece}{\end{center}}
\date{xxx 2013}

\usepackage[width=18.46cm,height=10in]{geometry}
\usepackage[noadjust]{cite}
\usepackage{srcltx}
\usepackage{psfrag}
\usepackage{epsfig}
\usepackage{amsmath}

\begin{document}

\title{Log-Periodic Dipole Array Antenna as a Chipless Radio-Frequeny Identification (RFID) Tag}

\author{S. Gupta, G. J. Li, R. C. Roberts and L. J. Jiang}

\abstract{A passive chipless Radio-frequency identification (RFID) tag based on log-periodic (LP) dipole array is proposed, where the tailorable band-rejection property of the LP aperture is utilized to realize large number of codes. The proposed tag principle is successfully validated using measurements, where the absence and presence of the band-rejection, is shown to carry the bit information. Its fabrication simplicity is also demonstrated by its implementation on a flexible substrate. Finally, two different tag formation schemes, based on specific set of resonance suppressions, are discussed in detailed. 
}

\maketitle

\section{Introduction}

Radio-frequency identification (RFID) systems have found diverse applications in the field of communications, ticketing, transportation, logistics, tracking, inventory, human identification, and security, to name a few \cite{Hartmann-SAW-UltrasonicSymp}. A typical RFID system comprises an interrogator (also called a reader) and many tags (also called labels). Recently, there is a strong research interest in passive chipless RFID tags, where the absence of the power supply and the integrated circuits (ICs), has shown promise for low-cost RFID solutions \cite{Preradovic-RFID-MicroMagazine}. Chipless tags, in particular, are useful under extreme environments such as extremely high or low temperatures that are not suitable for ICs. However, they are typically restricted to relatively small distances from the reader and suffer from low number of bits that can be encoded \cite{Preradovic-RFID-MicroMagazine}\cite{Hartmann-SAW-UltrasonicSymp}\cite{Zhang-TDRefect-HDPConf}\cite{Zhang-TDRefect-CircuitPConf}\cite{Vemagiri-TDRefectCirc-MOTL}.

Recently, a log-periodic dipole antenna aperture is proposed to be used as a information coding element in a chipless tag \cite{Gupta_LPDA}, to realize large number of bits. This paper provides the experimental validation of the LP based tag along with detailed discussion on its tag properties, with a focus on its design flexibility.

\section{Log-Periodic Dipole Array (LPDA) Tag}

\begin{figure}[h]
\centering
\psfrag{a}[c][c][1]{frequency (GHz)}
\psfrag{b}[c][c][1]{$S_\text{11, diff}$ (dB)}
\psfrag{c}[c][c][0.7]{\shortstack{$N=9$, $\beta = 0.4$~rad, $\alpha = 1.2$~rad \\ $\tau=0.8$, $\sigma = \tau^{1/2}$, $R =  43.7$~mm}}
\psfrag{d}[c][c][1]{$\alpha$}
\psfrag{e}[c][c][1]{$\beta$}
\psfrag{f}[c][c][1]{$2R$}
\psfrag{x}[c][c][1]{$N$-dipole arms}
\includegraphics[width=\columnwidth]{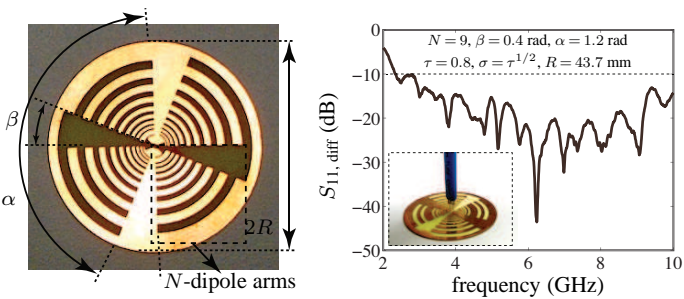}
\caption{Planar log-periodic (LP) antenna and its measured S-parameter response using a 100~$\Omega$ differential probe (shown in the inset). The physical parameters of the LP antenna are also shown in the figure.}\label{Fig:LP_Ap}
\end{figure}

A log-periodic (LP) antenna has impedance and radiation characteristics that are repetitive as a logarithmic function of frequency, resulting in a multi-octave bandwidth property \cite{Duhaml_LP}\cite{Carrel2}.  It has widespread applications in communications, electronic warfare systems from UHF to terahertz applications and UWB applications. An LP antenna consists of $N$ number of resonant dipoles whose dimensions are scaled by a constant parameter $\tau$, as shown in Fig.~\ref{Fig:LP_Ap}(a) along with the two angular parameters $\alpha$ and $\beta$ chosen for self-complimentary design, i.e. $\alpha + \beta = \pi$ \cite{Balanis-Antenna-Book}. The antenna is fed with a differential feed at the centre of the aperture and its typical S-parameter response is shown in Fig.~\ref{Fig:LP_Ap}(b), where a wideband matching response is seen, which is typical of LP antennas.

The discrete nature of resonant dipoles in an LP aperture can be used to encode information as was proposed in \cite{Gupta_LPDA}. Each dipole pair in the LP aperture is responsible for far-field radiation within a specific frequency band. The relation between two consecutive resonant frequencies of consecutive dipoles, is $f_{n-1}/f_n = \tau$, for any $n$, which can be used to determine the desired rejection bands. The proposed tag thus consists of an LP antenna aperture from which a specific combination of resonant dipoles are removed to introduce a band-rejection in the gain response of the antenna, as first demonstrated in \cite{Mruk}. By choosing a specific \textit{combination} of the resonant dipoles, the presence or absence of a null in the gain at a given frequency can be tailored, thereby realizing a specific RFID code.

\begin{figure}[h]
\centering
\includegraphics[width=\columnwidth]{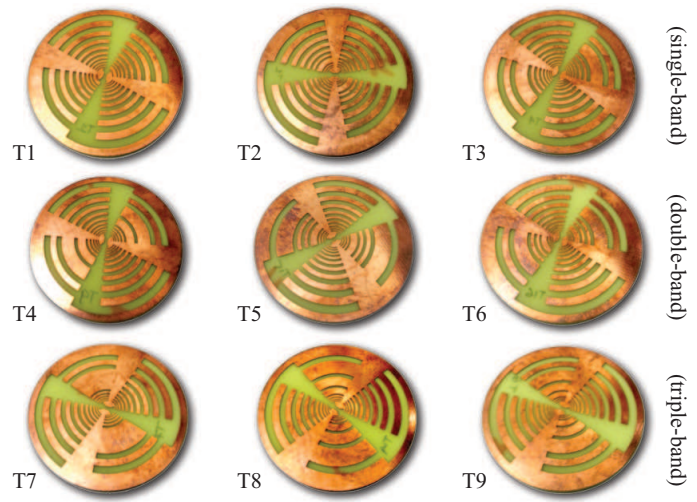}
\caption{Various fabricated prototypes of the printed log-periodic antenna with single, double or triple band-rejection response.}\label{Fig:LP}
\end{figure}

\begin{figure}[h]
\centering
\psfrag{a}[c][c][1]{frequency (GHz)}
\psfrag{b}[c][c][1]{S-parameters $S_\text{11, diff}$ (dB)}
\psfrag{c}[c][c][1]{\shortstack{$f_1$:   3.50 GHz\\
$f_2$:    4.10 GHz\\
$f_3$:   5.10 GHz\\
$f_4$:    6.20 GHz\\
$f_5$:   7.55 GHz\\
$f_6$:    9.25 GHz\\
$f_7$:    11.35 GHz}}
\includegraphics[width=0.75\columnwidth]{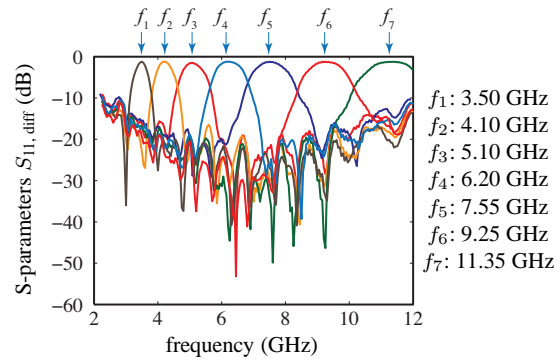}
\caption{Identification of centre frequencies of the rejection bands, corresponding to each dipole arm of the LP aperture. The consecuitve resonant frequencies closely follows the relation $f_{n-1}/f_n = \tau = 0.8\forall n$.}\label{Fig:KeyF}
\end{figure}

\begin{figure}[h]
\centering
\psfrag{a}[c][c][1]{frequency (GHz)}
\psfrag{b}[c][c][1]{S-parameters $S_\text{11, diff}$ (dB)}
\includegraphics[width=\columnwidth]{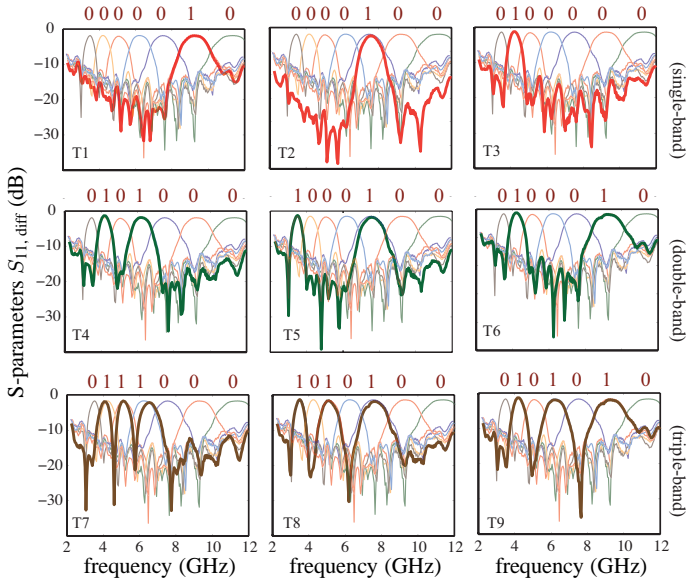}
\caption{Measured S-parameters of various LP tags of Fig.~\ref{Fig:LP}.}\label{Fig:Results}
\end{figure}

To validate this concept, various LP tags were fabricated, with some of prototypes shown in Fig.~\ref{Fig:LP}. The substrate used is FR4 with a $\varepsilon_r = 4.4$ and thickness of $0.8$~mm. The various tags differ from each other in the combination of band rejections. To characterize the specific frequency band associated with each resonant dipole, one set of tag was measured where, only one resonance is suppressed at a time (similar to first row of Fig.~\ref{Fig:LP}), ranging from the centre to the edge of the LP aperture. Fig.~\ref{Fig:KeyF} shows the corresponding S-parameters, where the key frequencies can be easily identified and associated with the corresponding dipole arm on the LP aperture. Now once those frequencies are known, their various combinations can be formed to realize specific RFID codes. Fig.~\ref{Fig:Results} shows the measured S-parameters of the various tags of Fig.~\ref{Fig:LP} where the presence and absence of a resonance is indicated as a binary bit. In each case, a distinct frequency response is clearly seen which successfully validates the proposed tag principle.

\section{Benefits and Features}

\begin{figure}[h]
\centering
\psfrag{a}[c][c][1]{frequency (GHz)}
\psfrag{b}[c][c][1]{S-parameters $S_\text{11, diff}$ (dB)}
\psfrag{c}[c][c][1]{$f_1$}
\psfrag{d}[c][c][1]{$f_2$}
\includegraphics[width=\columnwidth]{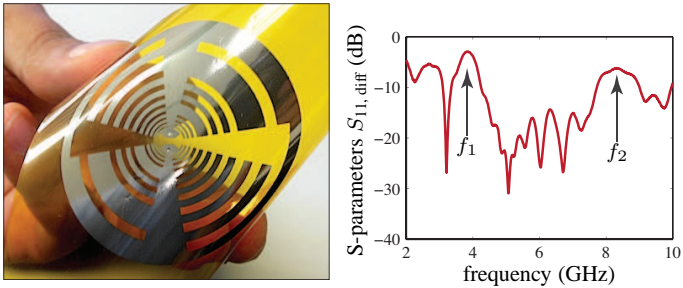}
\caption{Photograph of an LP tag antenna printed on a flexible substrate and its corresponding S-parameter response to illustrate the RFID code.}\label{Fig:flex}
\end{figure}

The proposed tag is a completely passive and chipless tag, thereby suitable in extreme environmental conditions, and compatible with planar PCB fabrication. Besides, it offers design simplicity where the aperture plays the dual role of efficient radiator and the coding element. To illustrate the fabrication flexibility of the proposed tag, an example tag was printed on a flexible substrate (DuPont Kapton HN) with $\varepsilon_r = 3.4$ and $25\mu$m thickness, as shown in Fig.~\ref{Fig:flex}. The corresponding S-parameters successfully demonstrates the code information.  

\begin{figure}[h]
\centering
\psfrag{f}[c][c][1]{frequency (GHz)}
\psfrag{a}[c][c][1]{VSWR ($Z_0 = 100$~$\Omega$)}
\includegraphics[width=\columnwidth]{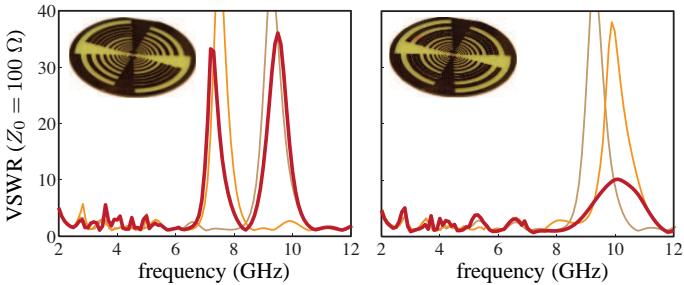}
\caption{Example of tag formation schemes and their typical VSWR responses obtained using FEM-HFSS.}\label{Fig:TagScheme}
\end{figure}

The number of possible code, for a given aperture size, depends on the number of radiating dipoles $N$, which can be controlled by choosing the growth parameter $\tau$. The tag formation can take two possible approaches, as shown in Fig.~\ref{Fig:TagScheme}: a) Suppressing the resonance located along the diagonals (two opposite quadrants) as is used in this paper, and b) Suppressing the resonances located in all four quadrants resulting in larger code combinations. In the first case, the total number of codes that can encoded is $2^N$. In the second case, it is found using full-wave simulations, that when two \textit{consecutive} resonances are suppressed, the two neighbouring resonances combine, leading to a poor VSWR as illustrated in Fig.~\ref{Fig:TagScheme}. Consequently not all combinations can be used and suppressing two consecutive resonances is thus not allowed. In this case, the total number of tag combinations form a Fibonacci number, i.e. $F_m = F_{m-1} + F_{m-2}$, with $F_0=0$ and $F_1=1$. For example, total combinations possible in an LP aperture with say $N=7$ dipoles, is $F_{14} = 377$.

Finally, the proposed tag is frequency scalable and intrinsically broadband  thereby capable of fast interrogation response. Moreover, the field-of-view of the LP antenna is large (typically larger than $120^\circ$) due to its dipole-like bi-directional radiation pattern, thereby making the tag suitable for interrogating from a bigger radiation space.

\section{Conclusion}
An experimental validation of a passive chipless RFID tag has been successfully shown using S-parameter characterization, both on a rigid and a flexible substrate to illustrate its design flexibility. The number of possible codes on a given LP aperture has been shown to depend on the choice of resonance suppression scheme, whereby in each case, a large number of code combinations can be achieved. The proposed solution is thus expected to provide chipless tag solution for large bandwidth RFID systems, by offering tag simplicity and large code combination. 

\vskip3pt
\ack{This work was supported by HK ITP/026/11LP, HK GRF 711511, HK GRF 713011, HK GRF 712612, and NSFC 61271158. Authors would also like to thank Mr. Zilong Ma for his help in measurements of various prototypes.}

\vskip5pt

\noindent S. Gupta, G. J. Li, R. C. Roberts and L. J. Jiang (\textit{Department of Electrical and Electronic Engineering, The University of Hong Kong, Hong Kong,  China.})
\vskip3pt

\noindent E-mail: shulabh@hku.hk

\bibliographystyle{aer}
\bibliography{ReferenceList}

%
%
%
%

\end{document}